\documentclass[preprint,prc,aps,epsfig]{revtex4}
\usepackage{psfig}
\usepackage{graphicx}
\def\beq{\begin{equation}}
\def\eeq{\end{equation}}
\def\beqa{\begin{eqnarray}}
\def\eeqa{\end{eqnarray}}

\begin{document}

\title{\sc Chromoelectric fields and quarkonium-hadron  interactions 
at  high energies}
\author{M.S. Kugeratski and F.S. Navarra} 
\affiliation{Instituto de F\'{\i}sica, Universidade de S\~{a}o Paulo\\
 C.P. 66318,  05315-970 S\~{a}o Paulo, SP, Brazil}

\begin{abstract}
We develop a simple model to study 
the heavy quarkonium-hadron cross section in the high energy limit. 
The hadron is represented by an external  electric color
field (capacitor) and the heavy quarkonium is represented by a small color 
dipole. Using high energy approximations we compute the relevant cross 
sections, which are then compared with results obtained with other methods.
\end{abstract}

\maketitle

\vspace{1cm}

\section{Introduction}

Quarkonium-hadron cross sections ($\sigma_{\Phi h}$) are a necessary tool to 
understand the forthcoming data on quarkonium production, which will become 
available at RHIC. In the last six years many efforts have been devoted to this 
problem \cite{nos1} and real progress has been achieved, especially in 
what concerns the cross sections at low energies, close to the dissociation
threshold. In the energy region far from threshold the situation is less clear 
and even the energy dependence is still subject of debate. Extrapolation from
calculations valid at low energies points to different directions.
Results obtained with the non-relativistic quark 
model \cite{wongs} indicate a rapidly falling cross section. This behavior is due to 
the gaussian tail of the quark wave functions used in the quark exchange model.   
This same behavior could be found within chiral meson 
Lagrangian approaches with the introduction of  $\sqrt{s}$ dependent form
factors \cite{osl}. In QCD sum rules  \cite{regras} the cross section was found to
be monotonically increasing with energy.

The calculations of $\sigma_{\Phi h}$  designed to be valid at high energies 
($\sqrt{s} \simeq 20$ GeV) are quite few: the Bhanot-Peskin (BP) approach 
\cite{pes,kar,kar2,arleo,oh,song},  perturbative QCD plus geometrical extrapolation
\cite{gerland}, the model of the stochastic vacuum (MSV) \cite{dnnr} and the light-cone
dipole formalism \cite{huf2000}. During the  
last years the leading order BP approach  has been used very often. However,   
the recent next to leading order calculations presented in \cite{song} show that, for 
the charmonium, the formalism breaks down because this system is not heavy enough. 
Most of the calculations mentioned above predict a rising cross section. 
In Ref. \cite{oh}, $\sigma_{\Phi h}$ falls with the energy and in Ref. \cite{dnnr} it 
stays constant.

If the quarkonium is treated as an ordinary hadron, its cross section 
for interaction with any other ordinary hadron must increase smoothly at higher 
energies, in much the same way as the proton-proton or pion-proton cross sections. 
The underlying reason is the increasing role
played by perturbative QCD dynamics and the manifestation of the partonic 
nature of all hadrons. However this partonic picture starts to be dominant only at
much higher energies ($\sqrt{s} \, \rangle \, 100$ GeV). In the energy region  
relevant for RHIC physics non-perturbative aspects are still very important. In the  
high energy calculations mentioned above, different non-perturbative ingredients were 
employed: moments of the gluon distribution in the hadron 
\cite{pes,kar,kar2,arleo,oh,song}; hadron and quarkonium wave functions 
\cite{gerland} and QCD vacuum expectation values (condensates) \cite{dnnr}. 

Since there are still discrepancies concerning numbers (which may vary by one 
order of 
magnitude for different estimates) and the energy behavior, we think that it is 
interesting to calculate $\sigma_{\Phi h}$ with a non-perturbative approach,  
putting
emphasis on the role played by the chromoelectric fields. In \cite{benzahra} a 
similar
treatment was adopted to study the quarkonium dissociation inside a QCD plasma. 
The 
color electric fields appearing in the transition matrix element were related to 
the 
color charge density of the medium, which, in turn, was computed in a specific 
model 
of the QGP. Here we start with a similar expression for the transition amplitude 
but, since we are in a purely hadronic phase, we must know the chromoelectric field 
inside nucleons and pions. There has been progress in the study of these fields, 
coming from models of the QCD vacuum \cite{michael}, 
from lattice QCD \cite{suga}, from the Field Correlator Method (FCM) \cite{simonov} 
and 
from Coulomb gauge QCD \cite{adam}. We hope that we can benefit from these advances and
use the profiles of the chromoelectric fields estimated in these works in our problem. 
For this purpose, we treat the interaction between  the quarkonium and hadron 
as being analogous to the interaction of a small dipole traversing a large capacitor 
and interacting with the color electric field but not with  its sources.  In the final 
part of this work we discuss the validity of  this last assumption. Using a contact 
interaction between a heavy quark (or antiquark) and a quark (or antiquark) we compute 
the corresponding cross section and find that it is indeed much smaller than the 
heavy quark-external field cross section. The model 
developed here bears some  resemblance to the Bhanot-Peskin  picture, but is much 
simpler. 
Some simplifying assumptions are used to  render the calculations quasi analytic and 
preserve the understanding of the basic physics.

\section{The model}

\subsection{The interaction Hamiltonian}

The starting point is the assumption that the quarkonium (dipole) is small compared 
with the hadron (capacitor). As a 
consequence, the $\overline{Q}-Q$ pair will interact mostly with the external color 
field but not with the 
(quark) sources.
Moreover,  the external color field is considered to have only low momentum components 
(``soft gluons'') and thus is able
to transfer only a small amount of energy, which will be barely  enough to dissociate 
the bound state. In the case of the charmonium, the typical binding 
energy is $\epsilon \simeq 0.6$ GeV. Therefore, in a first approximation
\beq
\epsilon \ll M_{\Phi}
\label{apr1}
\eeq
where $M_{\Phi}$ is the mass of the bound state ($M_{\Phi} \simeq 3$ GeV). In the case 
of the 
bottonium this approximation is even better. The binding energy is 
also small compared to the collision 
energy 
\beq
\epsilon \ll\sqrt{s}
\label{apr2}
\eeq

Inequality  (\ref{apr1})  justifies the use of quantum mechanical perturbation 
theory (the Born approximation) and  inequality 
(\ref{apr2})  justifies the use of the eikonal approximation, which, in this case, 
implies that the hadron
follows a straight line trajectory and remains essentially undisturbed during the 
interaction.  In Figure 1 we present our picture of the scattering and our choice of
coordinates, in the  quarkonium rest frame: $\vec{r}_{1}$ and  $\vec{r}_{2}$  
are the quark and antiquark coordinates  and  $\vec{E}^{a}$ is the 
chromoelectric field in the projectile, which will be a proton or a pion, moving with 
constant velocity $\vec{v}$ at impact parameter $\vec{b}$.

With these assumptions we can write the interaction Hamiltonian as:
\begin{equation}
H_{int} = g (T^{a}_{1} \vec{E}^{a}_{1}  \vec{r}_{1} + 
\overline{T}^{b}_{2} \vec{E}^{b}_{2} \vec{r}_{2}  )
\label{ham1}
\end{equation}
where $T^{a}$ ($\overline{T}^{b}$) are the generators of color group SU(3) in the 
fundamental (conjugate) representation.  $\vec{E}^{a}_{1}$ and  $\vec{E}^{b}_{2}$ 
are the chromoelectric fields generated by the hadron in motion (capacitor) and 
``felt'' by quark and antiquark in the 
bound state respectively. They have to be Lorentz transformed to the quarkonium rest 
frame, bringing to our calculation a Lorentz 
gamma factor, which is the source of the energy dependence ($\sqrt{s}$) of our results.  
We shall for the moment neglect the magnetic component, since it does not do any 
work on the charges and thus is not effective in the energy transfer. Besides, the 
magnetic interaction is inversely proportional to the quark mass, being thus suppressed. 

We can represent this external field by:
\begin{eqnarray}
\vec{E}^{a} ( r_e , t)&=&\gamma \vec{E}^{a}_{0}  \,\, 
exp \left(- \frac{(X -x_e)^2}{d^2}-\frac{(Y - y_e)^2}{d^2} \right) \,\,
 exp \left(- \gamma^2 \frac{\left[ v t - z_e \right]^2}{d^2} \right)
\label{pot}
\end{eqnarray}
with  $e = 1, 2$ . X, Y and Z are the hadron coordinates and $\gamma$ is the usual
Lorentz factor. $Z=vt$, because the 
hadron moves with velocity  $\vec {v}$ along the z axis. $\vec{E}^{a}_{0}$, which 
will be abreviated by  $E$, is the color electric field at the center of the 
projectile.  The projectile mean 
square radius is related to the parameter $d$ through:
$$
\sqrt{\langle r^2_{h} \rangle } \,  = \, 0.86  \,\, d
$$
We neglect the deflection of the hadron trajectory, 
because we are studying 
reactions in the high energy and non-perturbative regime, i.e., with low momentum 
transfer.  X and Y are related with the 
impact parameter $b$ by: $b^2 = X^2 + Y^2$.  
Notice that, by simplicity, we choose one preferencial direction for the field, 
in  this case, the $x$-axis.

Neglecting the CM motion, (\ref{ham1}) can be rewritten as
\begin{equation}
H_{int} = g (\frac{\lambda ^{a}}{2} E^{a}_{1}  + 
\frac{ \lambda ^{{b}^{T}}}{2} E^{b}_{2}  ) ( \frac {x_{1} -  x_{2}}{2}) 
\label{ham2}
\end{equation}
Also for the sake of simplicity, when working with (\ref{ham2}), we will take 
$x_{1} -  x_{2} \simeq a$, where $a$ is the typical separation between quark and 
antiquark.
Initially the quark-antiquark pair is in a localized  region of the space.

\subsection{The initial state}

The 
initial wave function of system has spatial and  color parts defined by:
\begin{equation}
\Psi _{i} =  f(r_1, r_2) c_n d_n
\label{psii}
\end{equation}
where  $c_n$ and $d_n$, with $n=1, 2, 3$, are the initial color vectors \cite{griff}
 for quark 
and antiquark respectively, taken in a  color singlet state. We choose
\begin{equation} 
f(r_1, r_2) = N_i \, exp[{ -\frac{\vec{r}_1 \, ^2}{{a}^2}}] \, 
exp[{-\frac{\vec{r}_2 \, ^2}{{a}^2}}] \,
exp ( - i \varepsilon _i t )
\label{psi}
\end{equation}
where $\varepsilon _i$ ($\varepsilon _i = M_{\Phi}$) is the quarkonium initial 
energy and $N_i$ is a normalization 
constant given by:
$$
N^2_i = (\frac {2}{\pi})^3 \, \frac{1}{a^6} 
$$
The initial wave function $\Psi _{i}$ describes the  confinement of quarks and also  
asymptotic freedom, as it  allows the quarks to be independent inside the bag. It is 
easy to see that the connection between the quarkonium  mean square radius and the 
parameter $a$ is
$$
\sqrt{\langle r^2_{Q \overline{Q}} \rangle } \,  =  \, 1.09 \,\,  a
$$

\subsection{The final state}

Under the action of the external field the initial wave function  $\Psi _{i}$ evolves 
to a final state $\Psi _{f}$:
\begin{equation}
\Psi _{f} =  t(r_1, r_2) c_j d_k
\label{psif}
\end{equation}
where $c_j$ and $d_k$, with $j, k = 1, 2, 3$, are the quark and antiquark final color 
vectors and $t(r_1, r_2)$ the spatial part of the wave function. In the final state of  
this reaction we have to deal with the transition of a pair of an excited quark and an 
antiquark to a pair of mesons $D$ - $\overline{D}$ (or $B - \overline{B}$).  This 
transition is highly 
non-perturbative and  has to be modelled.  We shall use here two approaches. \\

{\bf model A} \\

We first assume that the quark and antiquark are converted into two free mesons 
(a $M$ and a $\overline{M}$) which are thus described by plane waves:
\begin{eqnarray}
t_A(r_1, r_2) &=& N_{A} \,\, exp \, ( i \vec p_1 . \vec r_1 ) 
\,\,  exp \, ( i \vec p_2 . \vec r_2 ) \,\,
exp \, ( -i \varepsilon _f t) 
\label{inic1}
\end{eqnarray}
where $ \vec{p}_1$  and  $ \vec{p}_2$ are the meson momenta  and  $N_A$ is a 
normalization constant given by:
$$
N^2_{A} = \frac {1}{V^2}  
$$
with $V$ being an arbitrary normalization volume, which will be cancelled in the 
calculation of the cross section. In the above expression  
$\varepsilon _f$ is the final energy of 
the  $Q - \overline{Q}$ pair. The energy transferred during the reaction  must be 
sufficient to dissociate the 
bound state into a pair of  mesons with open  charm ($D \overline{D}$) or 
beauty ($B \overline{B}$) and therefore:
\beq
\varepsilon _f = \sqrt{(\vec{p}_1)^{2} + m_M^{2}} \,\, + \,\, 
                 \sqrt{(\vec{p}_2)^{2} + m_{\overline M}^{2}}
\label{epsf}
\eeq
where  $m_M$ ($m_{\overline M}$) is the mass of the meson coming out from the 
fragmentation of the quark (antiquark). With this definition of $\varepsilon _f$ we  
implicitly account for the conversion of quarks into hadrons, a process which  cannot 
be better described  in this simple model.

The assumptions (\ref{inic1}) and (\ref{epsf})  are reasonable but they   
represent a case of ''extreme freedom'' : they do not take into account the energy loss 
from a parent quark when it is converted to a (less energetic) final meson. This process  
is described, in certain situations,  by the fragmentation functions. Morevover, the  
final mesons can have any momentum and even though higher momenta will be naturally 
suppressed in the calculation, we are overestimating the phase space of the reaction. \\

{\bf model B} \\

Given these weak points of (\ref{inic1}) and (\ref{epsf}) we shall also use a second 
approach  for  the final state which is more conservative. We shall assume that the energy 
transferred to the heavy quarkonium $\Phi$ will transform it into an excited 
(but still bound) 
state $\Phi'$. The mass of this excited state will be taken to be slightly higher than the 
first charmonium and bottonium  excitations  $\Psi'$ and $\Upsilon'$ respectively. It is 
known that these excitations are very weakly bound. Therefore, by choosing slightly 
higher masses for them, which are above the $D$-$\overline{D}$ and $B$-$\overline{B}$
decay thresholds, we are simulating a fragmentation process to a pair of nearly at rest 
mesons.  This assumption is complementary to (\ref{inic1}) and (\ref{epsf}) since here 
we give to the heavy quarks only the ''minimal freedom''. The ground state wave function 
was chosen to be the Gaussian (\ref{psi}). Taking the harmonic oscillator as inspiration, 
we choose the wave function of the first excited state as a function which is odd in the  
$x$ direction (this is the direction of the chromoelectric field) and symmetric in $x_1$ 
and $x_2$:
\begin{eqnarray}
t_B(r_1, r_2) &=& N_{B} \, \frac{x_1 + x_2}{2} \,
exp[{ -\frac{\vec{r}_1 \, ^2}{{a'}^2}}] \, 
exp[{-\frac{\vec{r}_2 \, ^2}{{a'}^2}}] \,
exp ( - i \varepsilon _f t )
\label{inic2}
\end{eqnarray}
where the normalization constant is: 
$$
N^2_{B} = (\frac {4}{\pi})^3 \, \frac{1}{{a'}^8}
$$
and $a'$ is related to the size of the state $\Psi'$ or $\Upsilon'$.
Using the wave function (\ref{inic2}) has some advantages. In first place, it avoids 
the definition of a fragmentation mechanism with the introduction of new parameters. 
In second place, as it can be seen, (\ref{inic2}) is orthogonal to (\ref{inic1}), so 
that the matrix element $<\Psi_f|H_{int}|\Psi_i>$ is zero if the Hamiltonian is a 
constant. Notice that this does not happen when we use (\ref{inic1}) and therefore the 
approach A might contain spurious contributions. The same comment is valid for the  
calculations made in Ref. \cite{benzahra}. This makes the contrast between approaches 
A and B even more necessary. Finally, in what follows we shall 
use the Hamiltonian (\ref{ham1}), without the approximation $x_1-x_2\simeq a$ made in 
model A. \\

{\bf Transition amplitudes and cross sections} \\

The transition amplitude for model A  can be easily computed from (\ref{ham2}), (\ref{pot}), 
(\ref{psii}), (\ref{psif}) and (\ref{inic1}):
\begin{equation}
T_{fi} = \langle\Psi_f | H_{int} | \Psi_{i}\rangle = \int  dt \int d^3 \vec{r}_1   \int d^3  
\vec{r}_2  \,\,
\Psi^{*}_f (\vec{r}_1,\vec{r}_2)  H_{int}(\vec{r}_1,\vec{r}_2) 
\Psi_{i}(\vec{r}_1,\vec{r}_2)   
\label{tfi}
\end{equation}
An analogous expression holds for model B with the use of (\ref{ham1}), (\ref{pot}),
(\ref{psii}), (\ref{psif}) and (\ref{inic2}). 
We next take the amplitude squared  $|T_{fi}|^{2}  = T_{fi}^{*} \, T_{fi}$ and  
since color is not observed, we take the average of the all initial color  states  
and  the sum of all final states:
\begin{equation} 
|T_{fi}|^{2} \to  \overline {|T_{fi}|^{2}} \equiv \frac {1}{3} \sum _{n}  
\frac{1}{8} \sum _{a}
 \sum _{j} \sum _{k} |T_{fi}|^{2} 
\label{somacor}
\end{equation}
The cross section with model A is given by:
\begin{equation}
\sigma_A =  \int \frac{V}{(2\pi^3)} d^3 p_1  \, \int \frac{V}{(2\pi^3)} d^3 p_2 \, 
2 \pi \int _{0}^{\infty }db \, b \, \overline{|T_{fi}|^{2}}
\label{crossA}
\end{equation}

The above expression  is very simple and can be calculated almost  
analytically. Because of the gaussian Ansatz (\ref{pot}) and (\ref{psii}) we can 
easily integrate (\ref{tfi}) over the coordinates and over the impact parameter. In 
the last step of (\ref{crossA}), the integration over the phase space had to be done 
numerically. In \cite{fkn} we made the additional assumption that the outgoing mesons
are nearly at rest and we could thus simplify (\ref{epsf}) and perform the integration
over $\vec{p}_1$ and $\vec{p}_2$ analytically. Here we prefer to be more  ``exact''
and perform the last integrations numerically.

The cross section with model B is simply given by:
\begin{equation}
\sigma_B =  2 \pi \int _{0}^{\infty }db \, b \, \overline{|T_{fi}|^{2}}
\label{crossB}
\end{equation}
which, after the proper substitutions and integrations yields:
\begin{equation}
\sigma_B= \frac{32}{3} \pi^5 \,\, \langle gE_0 \rangle ^2 \,\, 
\frac{\gamma^2}{\gamma^2 -1}  \,\,
\frac{d^{10} \, a'^8 \, a^{10}}{(a'^2+a^2)^5  [a'^2 a^2  + d^2 (a'^2+a^2)]^3} \,\,   
exp{\Big(-\omega^2   \frac{\frac{\gamma^2 a'^2 a^2}{(a'^2+a^2)} + d^2  }
{2 (\gamma^2 -1) } \Big)}
\label{crossB2}
\end{equation}
where:
\begin{equation}
\omega = \varepsilon _f - \varepsilon _i \,\, = \,\,  M_{\Phi'} -  M_{\Phi}
\end{equation}

From the above expression we can observe that the cross section rises with the energy 
($\gamma$) and saturates at a constant value. The enhancement of the chromoelectric field 
is tamed by the Lorentz contraction of the projectile. As for the size parameters, $a$,
$a'$ and $d$, the cross section first rises and then falls with increasing values of the  
parameters. The values of the maxima strongly depend on the model and might change for a  
different choice of wave functions. However,  the physical picture is very simple. Expression 
(\ref{crossB2}) tells us that the probability of converting a quarkonium of  given initial 
size $a$ to a final state with size $a'$ tends to zero if  $a'=0$ or if  
$a'\to \infty$ because the overlap between these very different states and the 
initial state is zero. For the same 
reason the cross section vanishes for $a=0$ and for $a\to \infty$. The parameter $d$ 
is associated with the extension of the capacitor. When it goes to infinity the spatial  
dependence of the potential disappears, it becomes a constant and then  
$<\Psi_f|H_{int}|\Psi_i>  \to <\Psi_f|\Psi_i> = 0$.

\subsection{The interaction with the sources}

In the introduction it was assumed that the quarkonium is well  represented by a small 
dipole, which traverses a large capacitor. However this may be a too strong assumption 
because the dipole is not always so small. For example, comparing the size of the charmonium
with the size of pion we have tipically $\frac{a}{d} \simeq \frac{0.4}{0.6} \simeq 0.67$. 
Therefore it is necessay to include the interaction between the quark and antiquark in the
quarkonium with the sources (the ''plates'' of the capacitor) which may be either a quark 
and an antiquark in the case of the pion or quark and a diquark in the case of the proton.

In order to take these interactions into account we shall assume that the interaction between 
a quark (or diquark) in the capacitor and a charm quark (or antiquark) in the dipole can be
divided into a short distance and a long distance part. The later was already included before 
in the interaction with the chromoelectric fields produced by the sources. The former will be 
modelled as follows. \\

{\bf model C} \\

The short distance 
interaction can be approximated by the contact interaction part 
(the one with the delta function) of the one-gluon exchange potential \cite{brac}:
\begin{equation}
H_{int} = V_{OGE} =  \sum_{i=a,b} \sum_{j=1,2} 
\frac{\alpha_s}{4} \vec{\lambda_{i}}\cdot \vec{\lambda_{j}} 
\Big(\frac{1}{r_{ij}} \, - \,  \frac{2\pi}{3m_i m_j} 
\vec{\sigma_{i}}\cdot \vec{\sigma_{j}}  \delta^3(\vec{r_{ij}}) \Big)
\label{oge}
\end{equation}
where $\lambda$ and $\sigma$ are the Gell-Mann and Pauli matrices respectively, which are 
responsible for color and spin interactions. The Coulomb term in the above expression will
be neglected because it is of long range.  The labels $i=a,b$ and $j=1,2$ refer to 
particles in the capacitor and dipole respectively. With this notation, in the interaction 
between particle $a$ and $1$ the delta function above takes the form:
\begin{equation}
\delta^3(\vec{r_a}-\vec{r_1})=\delta(x_a-x_1) \times \delta(y_a-y_1)  
\times \delta(z_a-z_1)
\end{equation}
where $\vec{r_1}=(x_1,y_1,z_1)$ is the same as before and $\vec{r_a}=(x_a,y_a,z_a)$   
is the coordinate of the particle $a$ in the quarkonium rest frame.
In order to compute  the transition amplitude we need to 
know the new wave functions, which now include both the quarkonium and the capacitor. They 
are:

\begin{equation}
\Psi_i =f(\vec{r_1}, \vec{r_2})g(\vec{r_a}, \vec{r_b}) 
c_n d_n e_m h_m 
\label{psii_oge}
\end{equation}
and 
\begin{equation}
\Psi_f = t_C(\vec{r_1}, \vec{r_2})g(\vec{r_a}, \vec{r_b}) 
c_id_j e_lh_k 
\label{psif_oge}
\end{equation}
In the above expression the function $f$ is the same as before and given by (\ref{psi}).
The function  $t_C$ represents the spatial distribution of the heavy quarks in the final 
state, which is assumed to be an excited but still bound state, very much like in model B. 
However, if we would choose $t_C = t_B$, the transition amplitude 
$<\Psi_f|H_{int}|\Psi_i>$  would vanish  because the contact interaction does not depend on
the coordinates and hence $<\Psi_f|\Psi_i>$ is the  product of an odd by an even function of 
$x$,  
being thus zero. Since we are mostly interested in knowing the order of magnitude of this 
contact interaction we shall approximate the final state wave function by a gaussian, given 
by:
\begin{equation}
t_C(\vec{r_1}, \vec{r_2}) = N_C  \,  exp{(\frac{-r^ 2 _1}{a^ {'2}})} \,
exp{(\frac{-r^ 2 _2}{a^ {'2}})} \, e^{-i\varepsilon _f t}\,  
\label{t_oge}
\end{equation}
with the normalization constant given by:
\begin{equation}
N_C^2=\Big(\frac{2}{\pi}\Big)^3 \frac{1}{a'^6}
\end{equation}
The computation of the contact interaction requires the knowledge of the positions of the 
quarks in the capacitor, which is given by the function $g$
\begin{eqnarray}
g(\vec{r_b}, \vec{r_b}) &=& N_{P} \, exp{\Big[\frac{-(x_a-X)^2}{d^2}\Big]} 
exp{\Big[\frac{-(y_a-Y)^2}{d^2}
\Big]}  \nonumber \\
&\times& exp{\Big[\frac{-(x_b-X)^2}{d^2}\Big]} exp{\Big[\frac{-(y_b-Y)^2}{d^2}\Big]} 
\nonumber  \\
&\times& exp{\Big[\frac{-\gamma^2(z_a-Z)^2}{d^2}\Big]} 
exp{\Big[\frac{-\gamma^2(z_b-Z)^2}{d^2}\Big]}
\label{wavecap}
\end{eqnarray}
where $Z= v t$, $d$ and $\gamma$ have the same meaning as before and $N_P$ is the 
normalization constant of the projectile wave function, given by:
\begin{equation}
N_{P}^2=\frac{8\gamma^2}{\pi^3 d^6}
\label{norcap}
\end{equation}

Notice that $g$ is the same in the initial and in the final state. This assumption is 
consistent with the eikonal approximation introduced above and avoids the introduction 
of new parameters.

With these ingredients we can evaluate the transition amplitude:
\begin{eqnarray}
T_{fi} &=& \langle\Psi_f | H_{int} | \Psi_{i}\rangle  \nonumber  \\ 
&=& \int  dt \int d^3 \vec{r}_1   d^3  \vec{r}_2  
\int d^3 \vec{r_a}  d^3  \vec{r_b}  \,\,
\Psi^{*}_f (\vec{r}_1,\vec{r}_2,\vec{r_a},\vec{r_b}) \,
H_{int}(\vec{r}_1,\vec{r}_2, \vec{r_a},\vec{r_b})  \,
\Psi_{i}(\vec{r}_1,\vec{r}_2, \vec{r_a},\vec{r_b})   
\label{tfi_oge}
\end{eqnarray}
and the cross section:
\begin{eqnarray}
\sigma_C &=& \frac{2^{10}}{3^4} \pi \alpha_s ^2   \,\,
\Big(\frac{1}{m_a m_1}+\frac{1}{m_a m_2}+\frac{1}{m_b m_1}+\frac{1}{m_b m_2}\Big)^2 
 \nonumber \\
 &\times& \,  \frac{\gamma^2}{\gamma^2 - 1} \,\, 
\frac{a^6 a'^6}{(a'^2+a^2)^5 [d^2 (a'^2+a^2) + 2 a^2 a'^2]} \,\,
exp{\Big(-\omega^2   \frac{\frac{\gamma^2 a'^2 a^2}{(a'^2+a^2)} + d^2  }
{4 (\gamma^2 -1) } \Big)}
\label{crossC}
\end{eqnarray}
where we have used (\ref{somacor}) and the analogous expression for the sum and 
average over spins. Apart from a numerical factor,  (\ref{crossB2}) and (\ref{crossC}) 
have the same energy dependence. This is so because the same Lorentz contraction in the 
exponent of the Hamiltonian (\ref{ham1}) and (\ref{pot}) leading to (\ref{crossB2}) is now 
present in the capacitor wave function  (\ref{wavecap}). Moreover, the same Lorentz 
$\gamma$ factor, 
previously multiplying the $\vec{E}^{a}$ field  in (\ref{pot}) reappears now in the 
normalization constant (\ref{norcap}). The dependence of (\ref{crossC}) on $a$ and $a'$ is 
qualitatively the same as the one found in (\ref{crossB2}) and has the same physical origin. 
Finally, the cross section above is now a monotonically decreasing function of  $d$. 
The observed behavior with $d$ means that, in a larger  capacitor the quarks are
spread across a larger transverse area and it becomes more difficult for them to find the 
charm quarks in the target and  suffer a contact interaction.

\section{Results and discussion}

In the numerical estimates presented below, we shall adopt $d=0.8$ and $0.6$ fm for
the proton and the pion respectively. We shall also take $a=0.4$ and $0.2$ fm for the 
$J/\psi$ and $\Upsilon$ respectively and  $a'=0.8$ and $0.45$ fm for the $\Psi'$ and 
$\Upsilon'$. 
The bound states $\Psi$  ($m_{\Psi} = 3.07$ GeV) and $\Upsilon$ ($m_{\Upsilon} = 9.46$ GeV) 
will be, in model A, dissociated  into pairs of  mesons $D$ ($m_D=1.87$ GeV) and 
$B$ ($m_B=5.27$ GeV).  
The excited states used in models B and C have masses  $m_{\Phi'} = 3.8$ GeV and  
$m_{\Phi'} = 11$ GeV in the case of charmonium and bottonium respectively.
The value of the strong coupling constant and the constituent quark masses are the same 
used in \cite{brac}, i.e., $\alpha_s  = 0.64$, $m_q=0.3$ GeV, $m_c=1.2$ GeV, 
$m_b=4.74$ GeV and the diquark mass is $m_d=0.60$ GeV.

As it is clear from (\ref{ham2}) and (\ref{pot}),  we need to know the average 
value of the color electric field  in the projectile $g E \, = \, \langle h|gE|h\rangle $.
In a first approximation this number might be identified with the string tension
$\kappa \simeq 0.18 $ $GeV^2$ or $ \kappa \simeq 0.9$ $GeV/fm$. The string tension 
calculated in  \cite{adam} is somewhat larger. In \cite{simonov} the transverse 
profile of the string was studied. The strength of $\langle h|gE|h\rangle $ depends on the 
quark-antiquark 
(or quark-diquark) separation, being larger for larger systems and so far it has been 
calculated only for large systems. Therefore $\langle h|gE|h\rangle $ is another source of  
differences between a proton and a pion projectile.  Taking an average 
of the values found in \cite{simonov} we choose $\langle h|gE|h\rangle =1$ GeV/fm. 

As mentioned in the introduction, our model has common aspects with the BP approach. 
Therefore we shall, in what follows,  compare our results for $\sigma_{\Phi h}$ with 
those obtained by Kharzeev in \cite{kar2}:
\beq
\sigma_{\Phi h} = 2.5 \,\, \left( 1 \,\, - \,\, \frac{\lambda_0}{\lambda} 
\right)^{6.5} \,\, \mbox{mb}
\label{khar2}
\eeq
with $\lambda$ given by
\beq
\lambda \simeq \frac{(s-M^2_{\Phi})}{2 M_{\Phi}}
\label{lamb}
\eeq
and $\lambda_0 \simeq (M_h +  \varepsilon) $, where $M_h$ is the projectile mass and
\beq
\varepsilon = 2\,m_M - M_{\Phi} .
\label{enerliga}
\eeq

In Figure 2  we show the cross sections for the proton-charmonium dissociation
obtained with model A (dotted lines)  and model B (dashed lines) and compare them with 
the BP cross section (solid line with stars) given by (\ref{khar2}). The two upper curves 
are obtained with $\langle h|gE|h \rangle  = 1$ GeV/fm and the two lower curves with 
$\langle h|gE|h \rangle  = 0.57$ GeV/fm (model A) and 
$\langle h|gE|h \rangle  = 0.53$ GeV/fm (model B). With these smaller values of the 
chromoelectric field our curves come close to (\ref{khar2}).  
Figure 3 shows the corresponding cross sections for the  proton-bottonium dissociation.  
Again,  the two  upper curves are obtained with $\langle h|gE|h \rangle  = 1$ GeV/fm and  
the two lower curves with $\langle h|gE|h \rangle  = 0.69$ GeV/fm (model A) and 
$\langle h|gE|h \rangle  = 0.49$ GeV/fm (model B). As in the previous figure, reducing the 
value of $\langle h|gE|h \rangle$  leads to some agreement with (\ref{khar2}).
Given the conceptual resemblance between our model and the BP one, it is  
reassuring  to find a certain  similarity between the results, 
both in magnitude and energy behavior, once an appropriate value of  
$\langle h|gE|h\rangle $ is chosen.

In Figure 4 we show the cross section for $J/\psi$ dissociation by pions  
compared with results 
obtained the meson exchange model \cite{osl} (thin dotted line), with the quark exchange 
model \cite{wongs}(thin long dashed line), with short distance QCD (the BP approach)  
Eq. (\ref{khar2}) 
(thick solid line) and QCD sum rules \cite{nos1} (thin solid line). In spite of the fact 
that, at such low energies our approach looses validity, it is, nevertheless,  interesting 
to observe that 
our curve is in the center of the region  covered by the other calculations. 
In Figure 5  we compare the cross sections $p \, - \, J/\psi$ (upper curves) and 
$\pi \, - \, J/\psi$ (lower curves) calculated with models A (dotted lines) and B (dashed  
lines). In the 
high energy limit, where both cross sections are nearly constant, we observe that the 
relation between the cross sections is: 
\beq
\sigma_{p-\Phi } \, \simeq \, 3 \, \sigma_{\pi-\Phi } \,\,\,\,\,\,  \mbox{model A} 
\label{comp1}
\eeq
\beq
\sigma_{p-\Phi } \, \simeq \, 4.2 \, \sigma_{\pi-\Phi }  \,\,\,\,\,\,  \mbox{model B} 
\label{comp2}
\eeq
which in both cases is much larger than the one  expected from the additive quark model:
\beq
\sigma_{p-\Phi } \, \simeq  \frac{3}{2} \, \sigma_{\pi-\Phi } 
\label{aditive}
\eeq
This is remarkable since the additive quark model relation holds for other high 
energy scattering processes like  $\pi-p$ and $p-p$. Since $\langle h|gE|h\rangle $ was kept 
the same for both cases, this unexpected relation between the cross sections must come 
from differences in the wave functions. In Figure 6 we repeat this comparison for
the reactions  $p \, - \, \Upsilon$ and $\pi \, - \, \Upsilon$, finding (\ref{comp1}) for 
both models. We have kept  $\langle h|gE|h\rangle =1$ GeV/fm for both projectiles.  
Taking  $\langle p|gE|p\rangle > \langle \pi|gE|\pi\rangle$ would increase the
deviation from (\ref{aditive}).

In the high energy limit ordinary hadrons are expected to have  a geometrical  total 
cross section.  Since the quarkonium dissociation discussed here is a more specific 
reaction it is not obvious that its cross section follows a geometrical behavior. Such a
behavior was found in \cite{arleo}: $\sigma_{\Phi h} \propto \alpha_s a_0^2$ where $a_0$ 
is the Bohr radius of the quarkonium. In our case, as it can be seen from (\ref{crossB2}), 
(\ref{crossC}) and from the numerical evaluation of (\ref{crossA}),  we have a very 
non-trivial dependence on $a$. Since the initial state (containing the variable $a$) is the 
same, the difference between models comes from the spatial dependence of the final state. 
The plane waves in model A have no spatial scale. Therefore they are more  ''inclusive'' and 
so $\sigma_A$ should be closer to the quarkonium-hadron total cross section than  
$\sigma_B$. In 
model B the quarkonium ground state is converted into a resonance-like state, which wave 
function contains the size parameter of the resonance and distorts the final geometrical 
behavior. Therefore model A is closer to a geometrical behavior than model B.

In order to see how far we are from the geometrical behavior, we show in Figures 7 and 8
the dependence of $\sigma_A$ (dotted line) and $\sigma_B$ (dashed line) on $a$
for  charmonium (Fig. 7) and bottonium (Fig. 8) dissociation. The cross sections are divided 
by $a^2$ so that geometrical behavior translates into a horizontal line. 
We see that, whereas model A tends to this behavior, model B is far from a geometrical 
behavior. This indicates again that our  model is very sensitive to the choice of the final 
state wave functions.

In Figure 9 we show the cross section $\sigma_C$ (\ref{crossC}) for charmonium dissociation 
by protons (solid line) and by pions (dashed line). In Figure 10 we show the same quantitity 
for bottonium dissociation. We use the central values for $a$, $a'$, $d$ and $\alpha_s$. 
We can see that, in all processes, the cross sections are more than
two orders of magnitude smaller than the corresponding cross sections computed with model A 
or model B. No possible change in parameters could make these cross sections  comparable. 
Another  feature of these curves is that the cross sections for $J/\psi$ 
dissociation by pions are larger than those for protons by a factor close to $4$. This might be
guessed looking at (\ref{crossC}). The pion is light quark-antiquark system and the proton is 
light quark-diquark dipole. The diquark is twice heavier than a constituent quark.  
Whereas for the pion $m_b=m_a$, for the proton we have $m_b= 2 m_a$. 

Before concluding we would like to make a remark concerning medium effects on the cross 
sections calculated above. We are primarily studying  reactions which happen before 
thermalization (in nucleus-nucleus collisions) or with no thermalization at all 
(in proton-nucleus collisions). The formation time of the heavy quark pair is of the 
order of $0.2$ fm. The thermalization time of hadronic matter formed in heavy ion collisions 
is a model dependent quantity. Early estimates  pointed to $1$ fm. Recent estimates 
\cite{heinz} point to  $0.6$ fm.   Even taking seriously this last number, it is fair to 
say that heavy quark pair  production (and collision with a hadron at 
high energies) precedes the formation of an equilibrated medium.  After thermalization, the 
energy is completely redistributed and collisions occur at energies of the order of the 
temperature ($< 1 $ GeV).   In this regime we do not expect our approach to be valid.
The effects of a thermal medium on the heavy quarkonium are known \cite{wong}: the string 
tension becomes weaker, the quarkonium size increases and its mass decreases. These 
effects are, all of them, very small except if we get close to the deconfinement 
transition temperature. In view of these considerations we have neglected medium effects in
our calculations.

\section{Conclusions}

We have developed a simple model for the non-perturbative quarkonium-hadron interaction. At 
the present stage of the field,  this sort of model is still useful to organize the ideas. 
We 
tried to make simple and yet realistic choices for the interaction Hamiltonian and for the 
wave functions. In particular we have treated the final state in two very different and 
complementary ways. Simple models are not appropriate to provide very precise results but 
they can help in determining  the order of magnitude of the cross sections and their 
behavior  
with the reaction energy. Having said that, we can summarize our conclusions as follows. \\
i)  The charmonium-hadron cross section is of a few milibarns. The bottonium-hadron cross 
section is about four times smaller. This is in agreement with most of the previous 
calculations. \\
ii)  All cross sections grow with the reaction energy and reach a plateau in the high energy 
limit. This is in agreement with the BP approach. \\
iii)  In this limit they do not obey the simple relations derived from the additive quark 
model.\\  
iv)  Also in this limit our cross sections deviate significantly  from the geometrical 
behavior 
($\sigma \propto a^2$). \\
v) The contact interactions between the heavy quarks and the light quarks in the light 
hadrons 
is negligible compared to the long-distance quark - $\vec{E}^{a}$ field interaction. This is 
surprising since sometimes the dipole and the capacitor have similar sizes. This finding 
gives 
a posteriori support to our model and also to the BP approach.

Conclusion i) may be relevant  for RHIC and LHC physics. Conclusions ii, iii and iv suggest 
that the heavy quarkonium has interaction properties which are very different from light 
hadrons. This has been conjectured before. In particular, in \cite{dnw} this difference was 
attributed to the fact that in heavy quarkonia the energy is mainly stored in the masses 
whereas  in light hadrons the energy (mass of the hadron) comes mostly from the gluonic 
fields.

\vspace{1cm}
 
\underline{Acknowledgements}: 
This work has been supported by CNPq and FAPESP. 
We are indebted to M. Nielsen, D.A. Foga\c{c}a and F. Dur\~aes   for 
fruitful discussions.
\vspace{0.5cm}

\eject


\eject

\begin{figure} \label{fig1}
\centerline{\psfig{figure=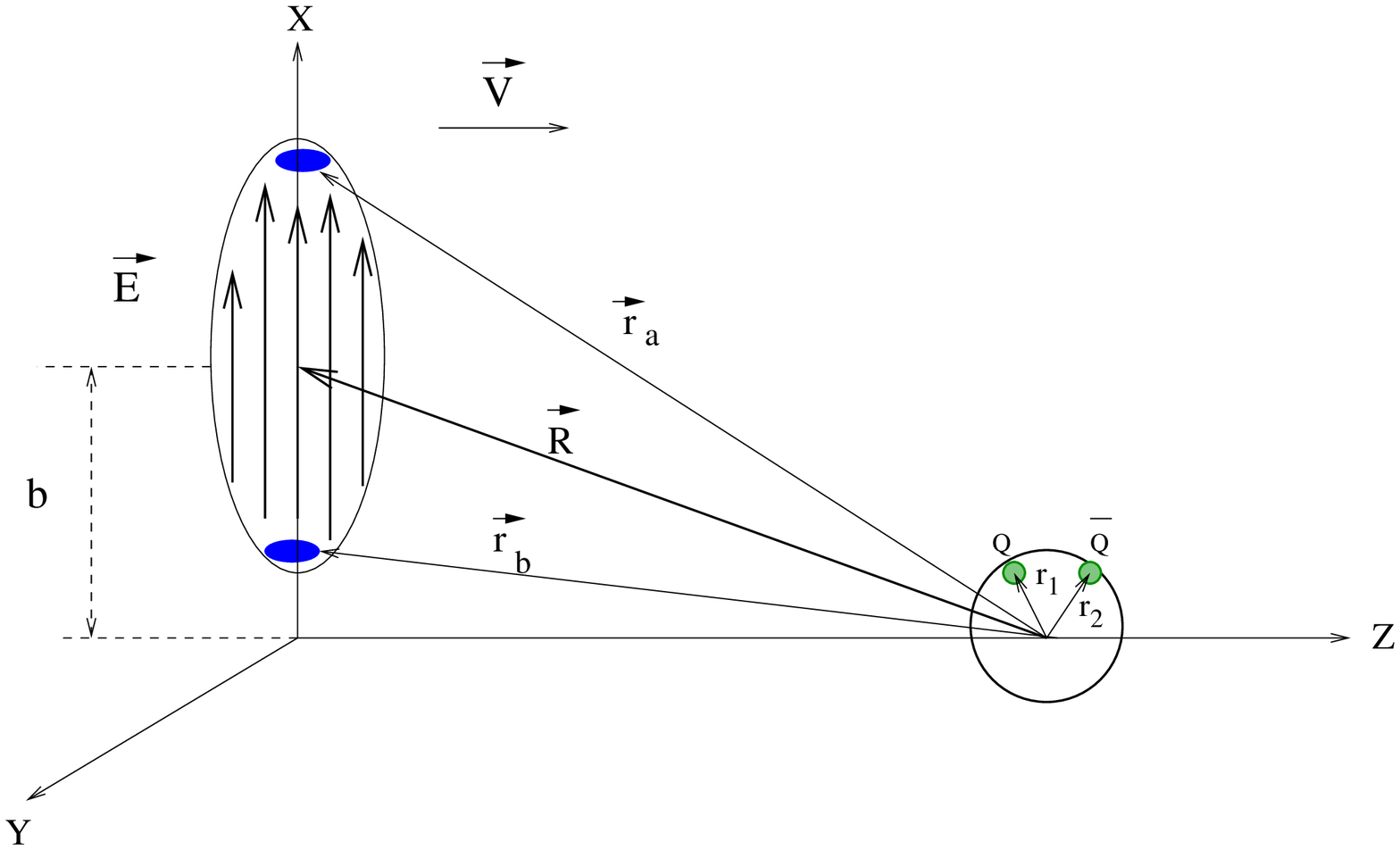,width=16cm,angle=0}}
\vskip -0.5cm
\caption{Quarkonium-hadron interaction in the quarkonium rest frame. The hadron is a  
capacitor moving to the right. The  $\vec{E}^{a}$ field is along the $x$ direction. The thick 
black dots are the capacitor ''plates'': quark-antiquark for the pion and quark-diquark for 
the proton.}
\end{figure}

\vskip 2.0cm

\begin{figure} \label{fig2}
\centerline{\psfig{figure=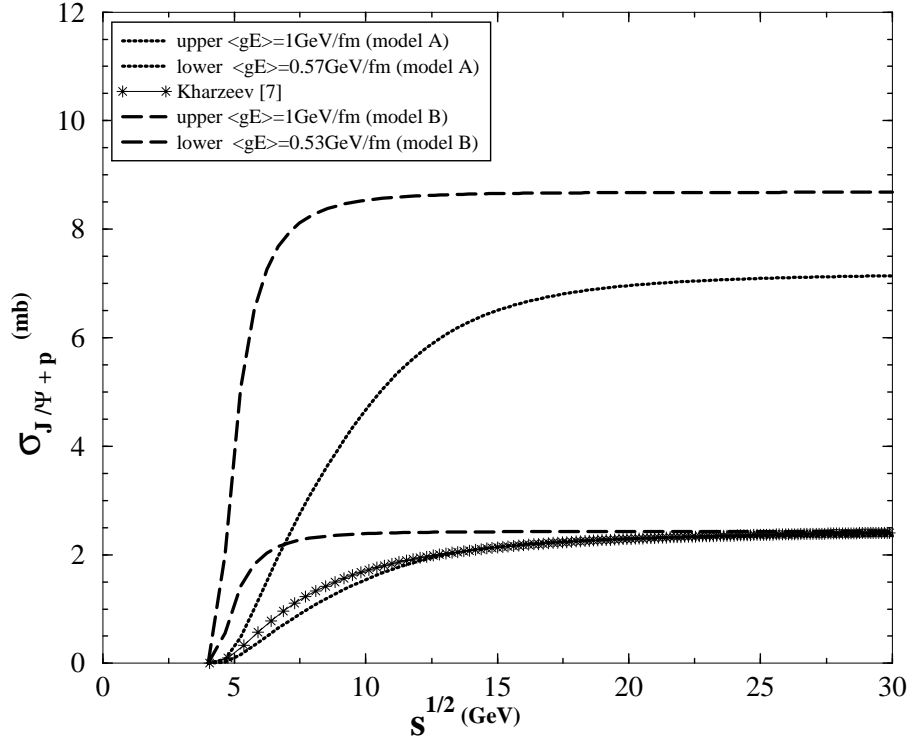,width=12cm,angle=0}}
\vskip -0.5cm
\caption{$J/\psi - p$ cross section with model A (dotted lines), model B (dashed lines) 
and with the Bhanot-Peskin approach (lines with stars). Upper curves: stronger $\vec{E}^{a}$ 
field. Lower curves: weaker $\vec{E}^{a}$ field.}
\end{figure}

\begin{figure} \label{fig3}
\centerline{\psfig{figure=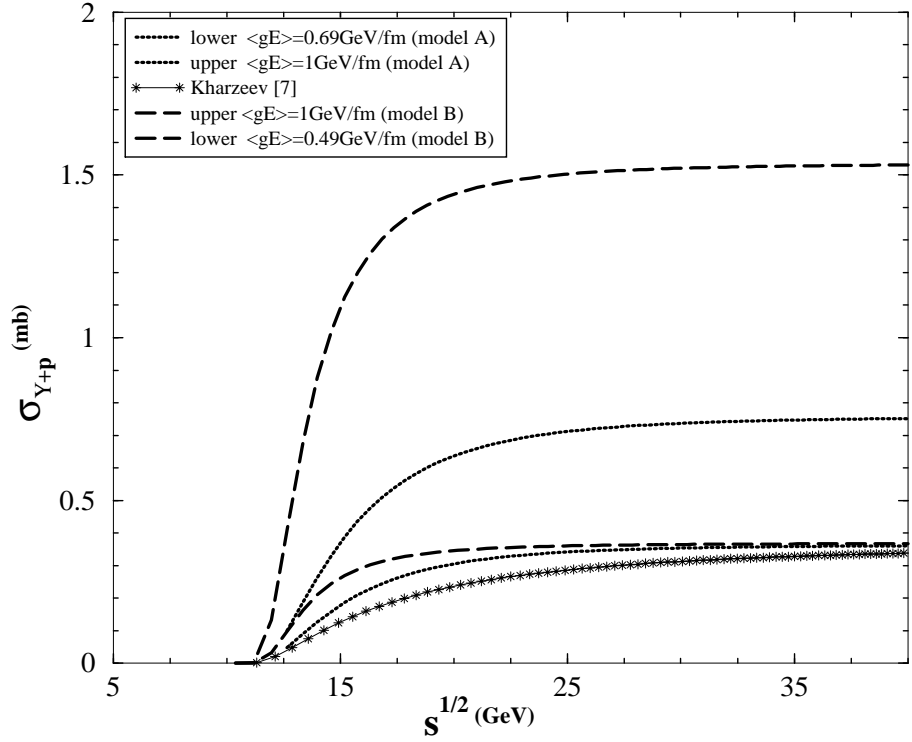,width=12cm,angle=0}}
\vskip -0.5cm
\caption{Same as Fig. 2 for the $\Upsilon + p$ cross section.}
\end{figure}


\begin{figure} \label{fig4}
\centerline{\psfig{figure=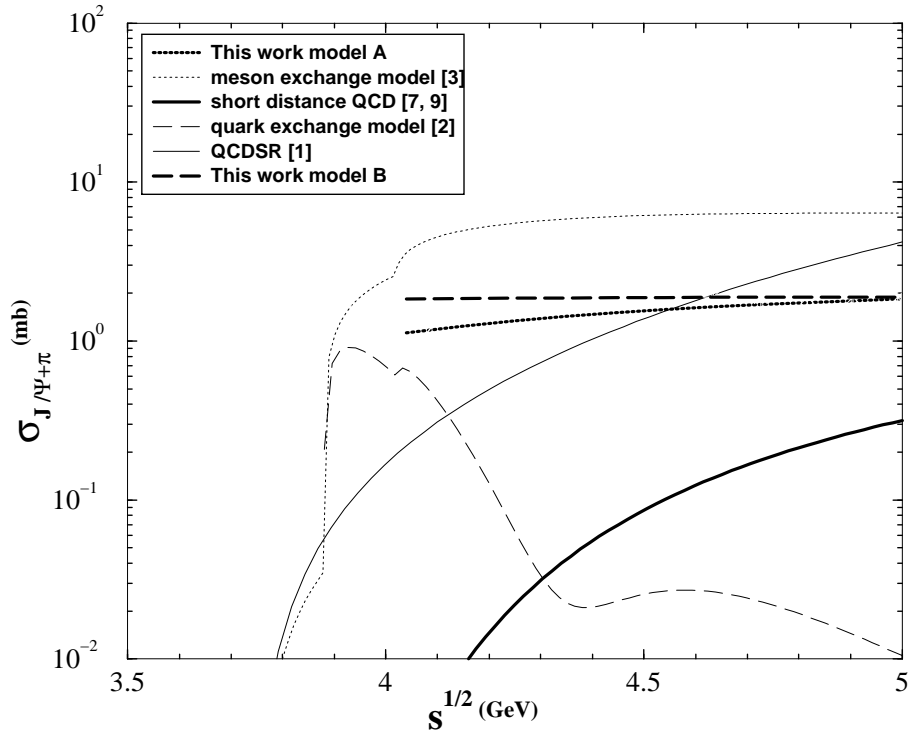,width=12cm,angle=0}}
\vskip -0.5cm
\caption{Pion-charmonium cross section as a function of  $\sqrt{s}$ with several models.}
\end{figure}

\begin{figure} \label{fig5}
\centerline{\psfig{figure=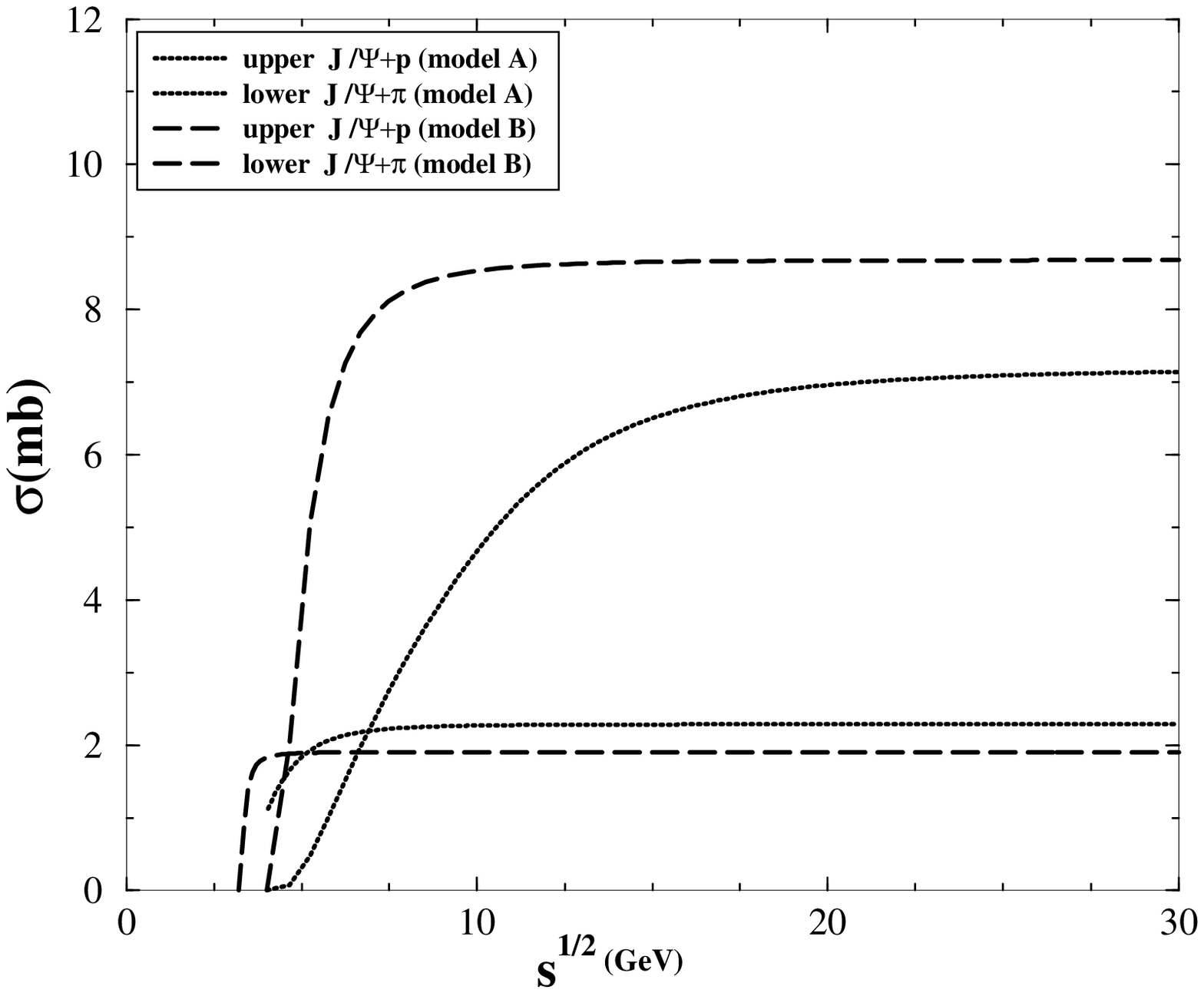,width=12cm,angle=0}}
\vskip -0.5cm
\caption{Charmonium-hadron cros section with model A (dotted lines) and model B (dashed lines). Upper curves: $\sigma_{J/\psi}-p$. Lower curves $\sigma_{J/\psi}- \pi$.}
\end{figure}

\begin{figure} \label{fig6}
\centerline{\psfig{figure=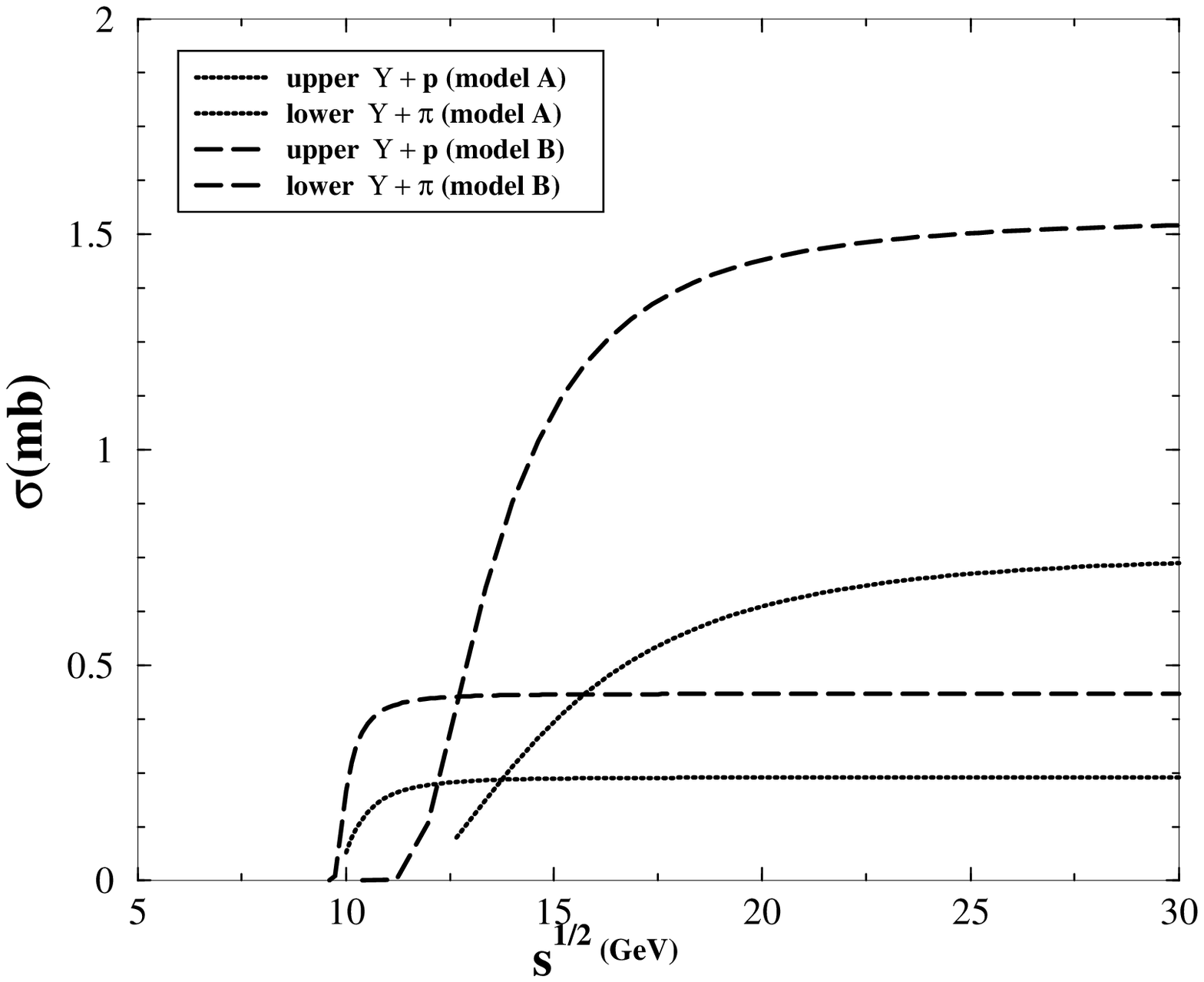,width=12cm,angle=0}}
\vskip -0.5cm
\caption{Same as Fig. 5 for  bottonium-hadron cross sections.}
\end{figure}



\begin{figure} \label{fig7}
\centerline{\psfig{figure=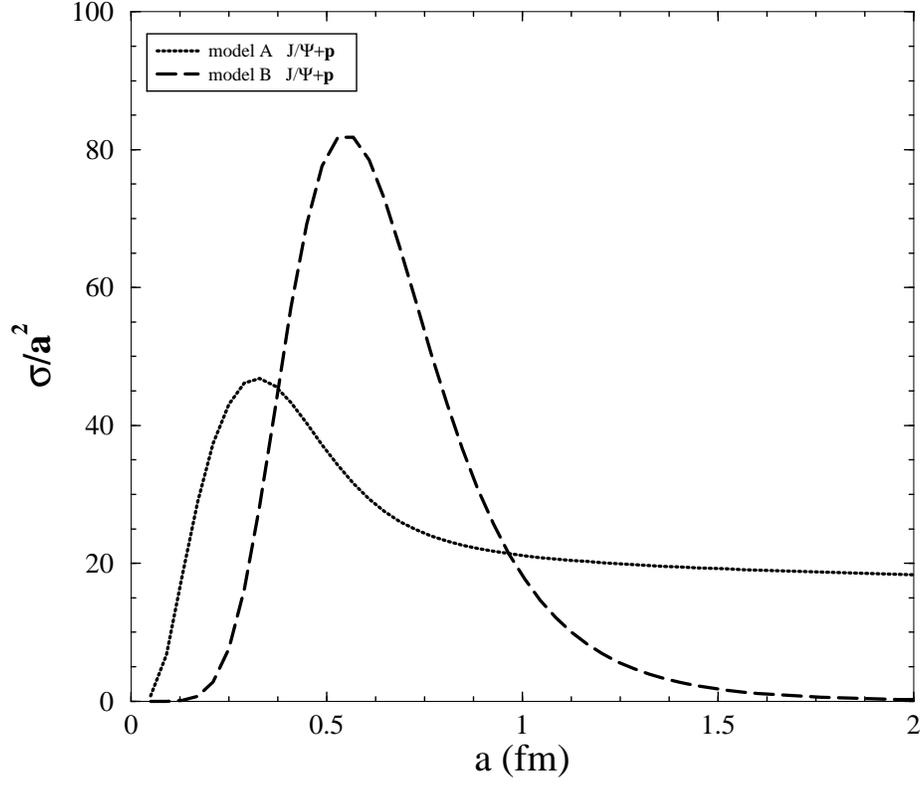,width=12cm,angle=0}}
\vskip -0.5cm
\caption{Charmonium-hadron cross sections as a function of the charmonium size parameter 
for model A (dotted line) and for model B (dashed line).}
\end{figure}

\begin{figure} \label{fig8}
\centerline{\psfig{figure=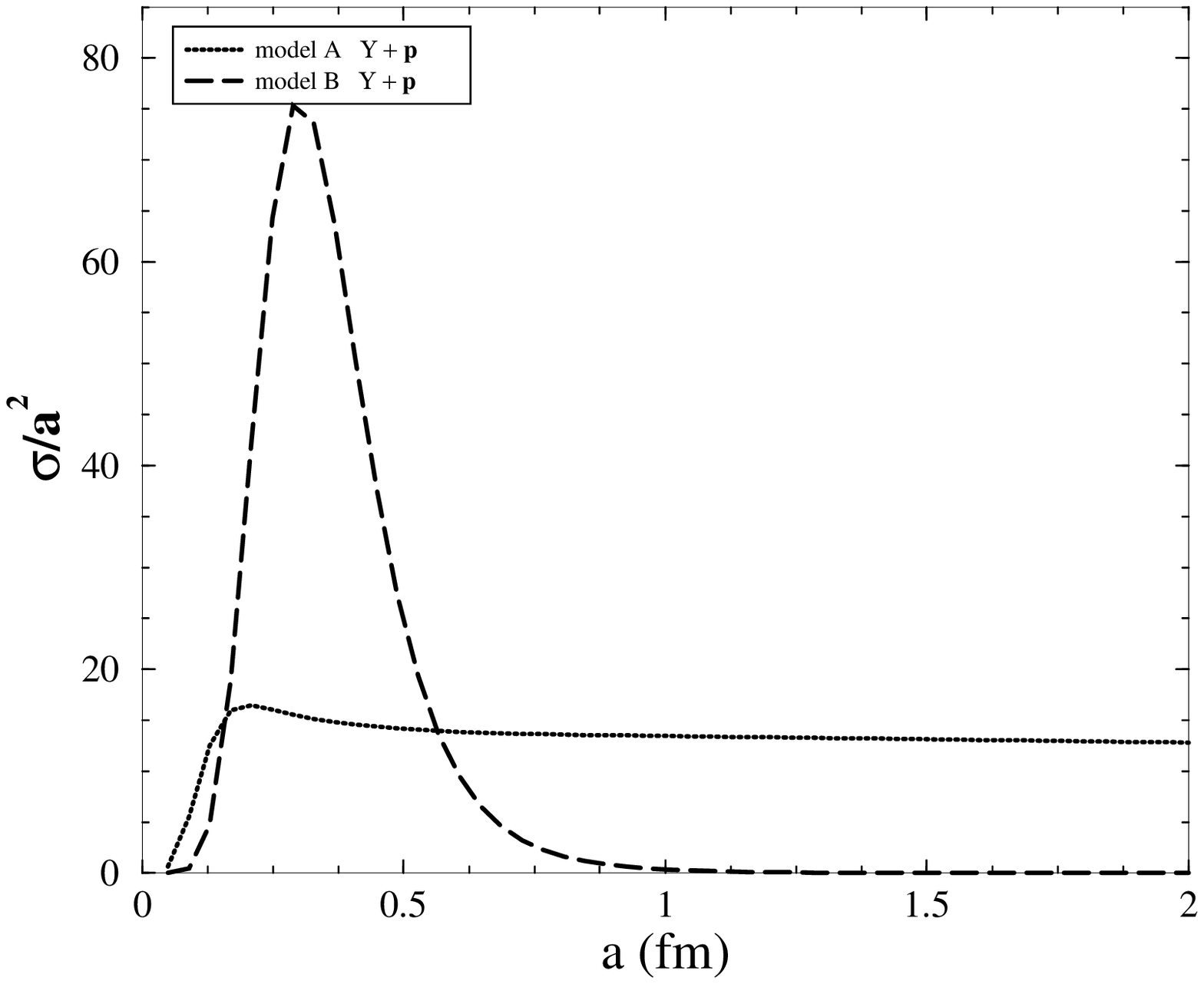,width=12cm,angle=0}}
\vskip -0.5cm
\caption{Same as Fig. 7 for bottonium-hadron cross sections.}
\end{figure}

\begin{figure} \label{fig9}
\centerline{\psfig{figure=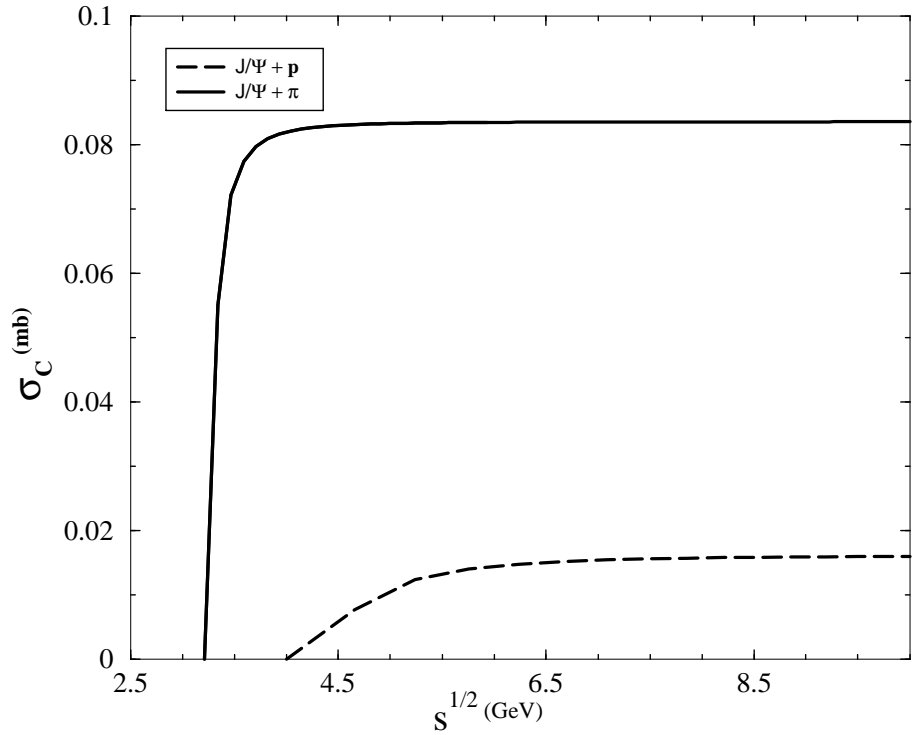,width=12cm,angle=0}}
\vskip -0.5cm
\caption{Charmonium-proton (dashed line) and charmonium-pion (solid line) cross sections 
calculated with model C (contact interactions).}
\end{figure}


\begin{figure} \label{fig10}
\centerline{\psfig{figure=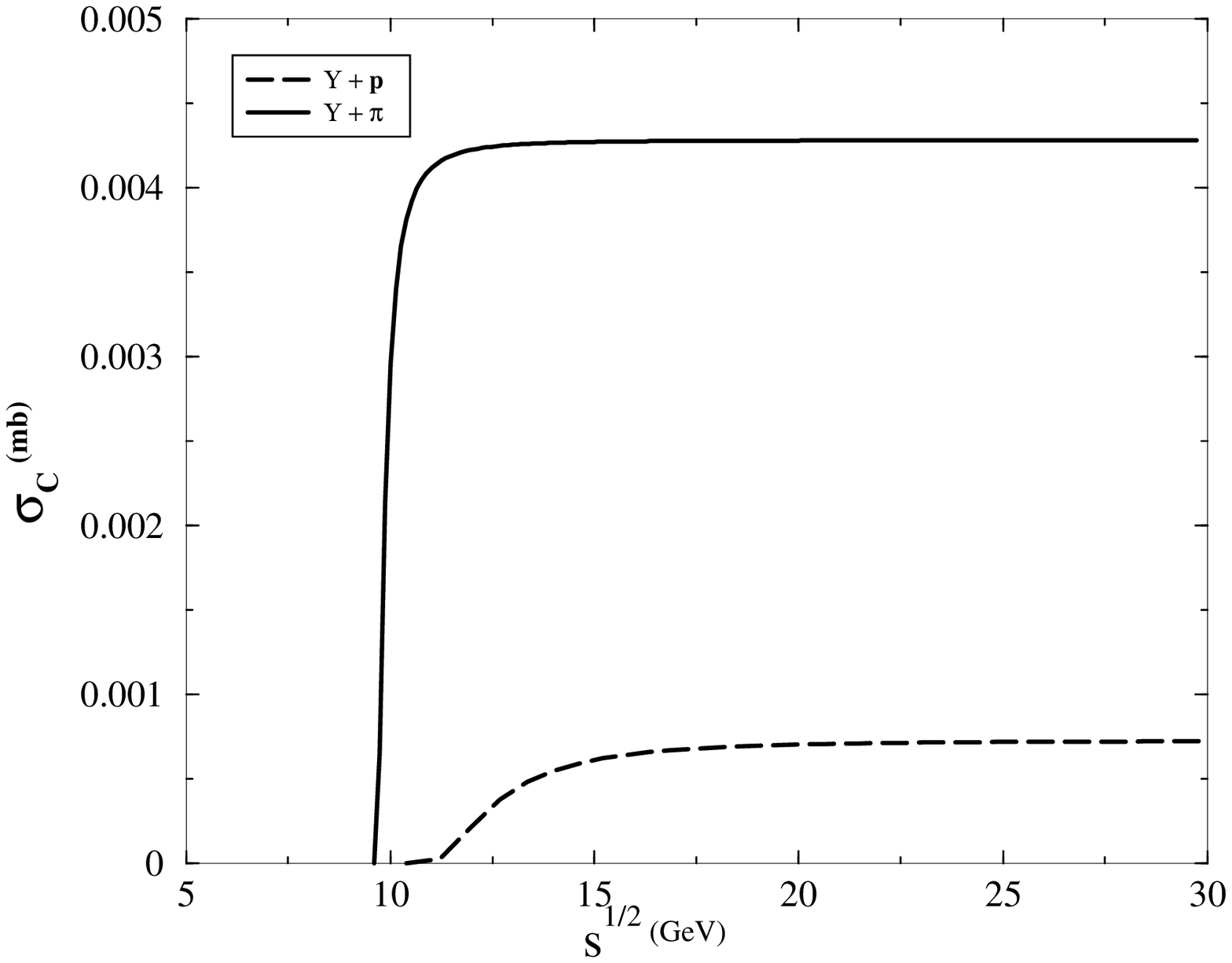,width=12cm,angle=0}}
\vskip -0.5cm
\caption{Same as Fig. 9 for bottonium-hadron  cross sections.}
\end{figure}

\end{document}